\documentclass[a4paper]{spie}  

\usepackage{amsmath,amsfonts,amssymb}
\usepackage{graphicx}
\usepackage{tabularx}
\usepackage{float}
\usepackage{booktabs}
\usepackage{subcaption}
\usepackage[colorlinks=true, allcolors=blue]{hyperref}
\usepackage{array}
\newcolumntype{Y}{>{\centering\arraybackslash}X}
\usepackage{comment}

\title{Testing of a 15-Positioner Module Based on the MPS Design for Stage-5 Telescopes}

\author[a]{Jonathan Wei}
\author[a]{Oliver Pineda Suárez}
\author[a]{Malak Galal}
\author[a]{Maxime Rombach}
\author[a]{Sébastien Pernecker}
\author[b]{Stefane Caseiro}
\author[b]{Corentin Magnenat}
\author[b]{Florian Boeckle}
\author[a]{Jean-Paul Kneib}

\affil[a]{Institute of Physics, Laboratory of Astrophysics, Ecole Polytechnique Federale de Lausanne (EPFL), Observatoire de Sauverny, CH-1290 Versoix, Switzerland}
\affil[b]{Micro Precision Systems AG (MPS), Bienne, Switzerland}

\authorinfo{Further author information: (Send correspondence to J.W. and O.P.S.)\\J.W.: E-mail: jonathan.wei@epfl.ch\\  O.P.S.: E-mail: oliver.pinedasuarez@epfl.ch}

\begin{document} 
\maketitle

\begin{abstract}
Assessing the performance metrics of theta-phi micro-robotic positioners is a key step toward understanding their operational behavior and ensuring that they meet the specifications of Stage-5 astronomical facilities, including the Chinese MUST, the American Spec-S5, and the European WST. A detailed examination of these metrics enables a clear evaluation of the system’s current capabilities while also revealing aspects that require further refinement, which in turn guides improvements in design and manufacturing. In this work, carried out in collaboration with Micro Precision Systems (MPS) in Switzerland, we present the results of positioning performance and angular tilt characterization conducted on a 6.2-mm-pitch robotic positioner module developed for high-density fiber placement in next-generation spectroscopic instruments. The prototype unit evaluated in this study was produced by MPS. The measured metrics, including positioning repeatability, datum repeatability, backlash, non-linearity, and angular tilt, are compared directly with the nominal performance targets defined for Stage-5 telescope systems.
\end{abstract}


\section{INTRODUCTION}
Robotic fiber-positioner systems are fundamental for large scale spectroscopic surveys using Multi-Object Spectrograph instruments. The demand for spectroscopic maps that acquire the spectra of larger galaxy populations at higher redshifts continues to grow\cite{mainieri_wide-field_2024}. Achieving these goals requires denser arrays of miniaturized fiber-positioning robots. This, in turn, introduces new engineering challenges.

\begin{figure}[H]
  \makebox[\textwidth][c]{%
    \begin{minipage}{0.7\textwidth}
      \centering
      \captionsetup[subfigure]{justification=centering}
      \begin{subfigure}[b]{0.5\textwidth}
        \centering
        \includegraphics[width=\linewidth]{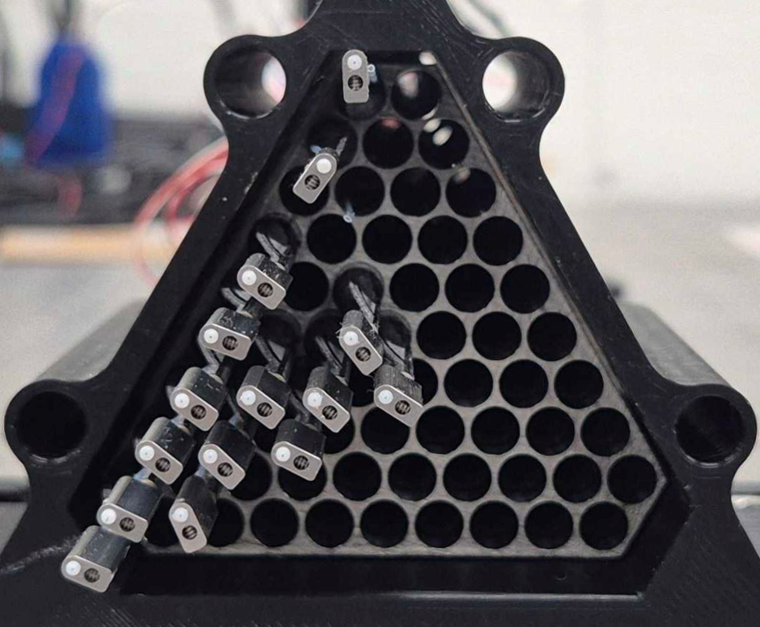}
        \caption{}
        \label{fig:protomps}
      \end{subfigure}
      \hspace{.3cm}
      \begin{subfigure}[b]{0.45\textwidth}
        \centering
        \includegraphics[width=\linewidth]{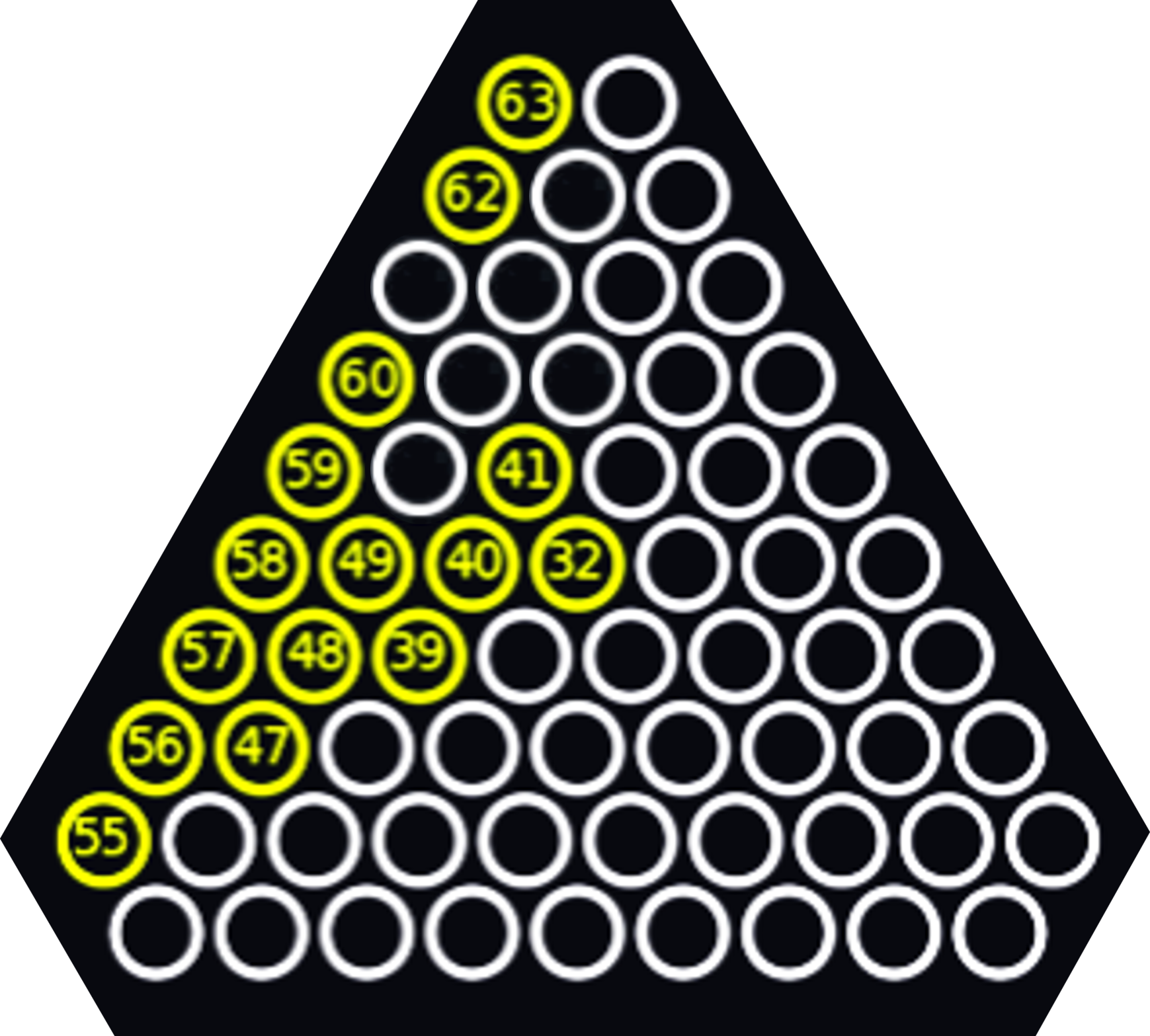}
        \caption{}
        \label{fig:protompsnum}
      \end{subfigure}
    \end{minipage}
  }
  \vspace{.1cm}
  \caption{(a) MPS 15-Fiber Positioner Prototype; (b) Numbering Convention for Fiber-Positioner Identification}
  \label{fig:numconv}
\end{figure}

Analyzing the positioning capabilities of these systems is important to evaluate design and manufacturing improvements and constraints. In this document the performance testing results obtained from the prototype designed and manufactured by the Swiss company Micro Precision Systems AG (MPS) are analyzed. Those performances can then be compared to the ones of the trillium module manufactured by Orbray \cite{pinedasuarez2026trillium}.

The prototype and the numbering convention used to identify each individual positioner are shown in Figure \ref{fig:numconv}. This modular design is based on the work proposed in Ref. \citenum{SilberModule} which address the design of this modular approach to address the scalability and Ref. \citenum{InvestigationsCoverage} which focuses on the analysis of an optimal amount of positioners per module considering the trade-off between coverage and focal surface fitting.

\section{METHODOLOGY}

This section outlines the experimental methodology employed to evaluate the performance of the fiber-positioners in this module.

\subsection{XY TEST}

The characterization of the fiber positioners with the XY test is based on the following metrics used to evaluate their behavior: repeatability, datum performance, backlash, and non-linearity. Taken together, these metrics provide a broad assessment of the system’s positioning performance.

Repeatability describes the capability of the system to achieve a commanded target position when approached multiple times from the same direction, either clockwise or counterclockwise. It is assessed through repeated positioning cycles and quantified from the dispersion of the measured positions relative to the commanded target. Small repeatability errors are indicative of precise and reliable motion, supporting accurate calibration and compliance with positioning requirements. The current repeatability tests have been conducted with movement of 30°.

Datum repeatability evaluates the consistency of the hard-stop reference used during homing. It is determined by repeatedly driving the mechanism to the datum position until the stop is engaged and recording the resulting fiber tip location. Since this reference is used in the kinematics model of the positioner, repeatable datum measurements are essential to ensure that the model is accurate.

Backlash represents the mechanical slack introduced by clearances in transmission components such as gears and couplings. It is characterized by measuring the positional offset obtained when the same target is approached from opposite directions. Reduced backlash is essential in order to have small approach moves to the target, when approaching every target from the same side. This allows to have to consider less margin when planning approach trajectories.

Non-linearity quantifies the extent to which the actual fiber tip displacement departs from the ideal linear response to commanded motion over the actuator range. It is determined by comparing measured centroid positions to their expected locations at multiple target points. These deviations, typically associated with gear transmission effects and mechanical tolerances, reflect the degree to which the commanded position deviates from the actual measured fiber position. While non-linearity can be compensated as long as it is repeatable, high non-linearity often points to bad mechanical tolerances in the gears and leads to slower convergence to target, partly due to higher variance in the non-linearity in those cases. High non-linearity variation also makes the output shaft speed fluctuate. Thus, lower non-linearity over the whole travel range of each arm of the positioner is desired.

\begin{figure}[H]
    \centering
    \includegraphics[width=0.4\linewidth]{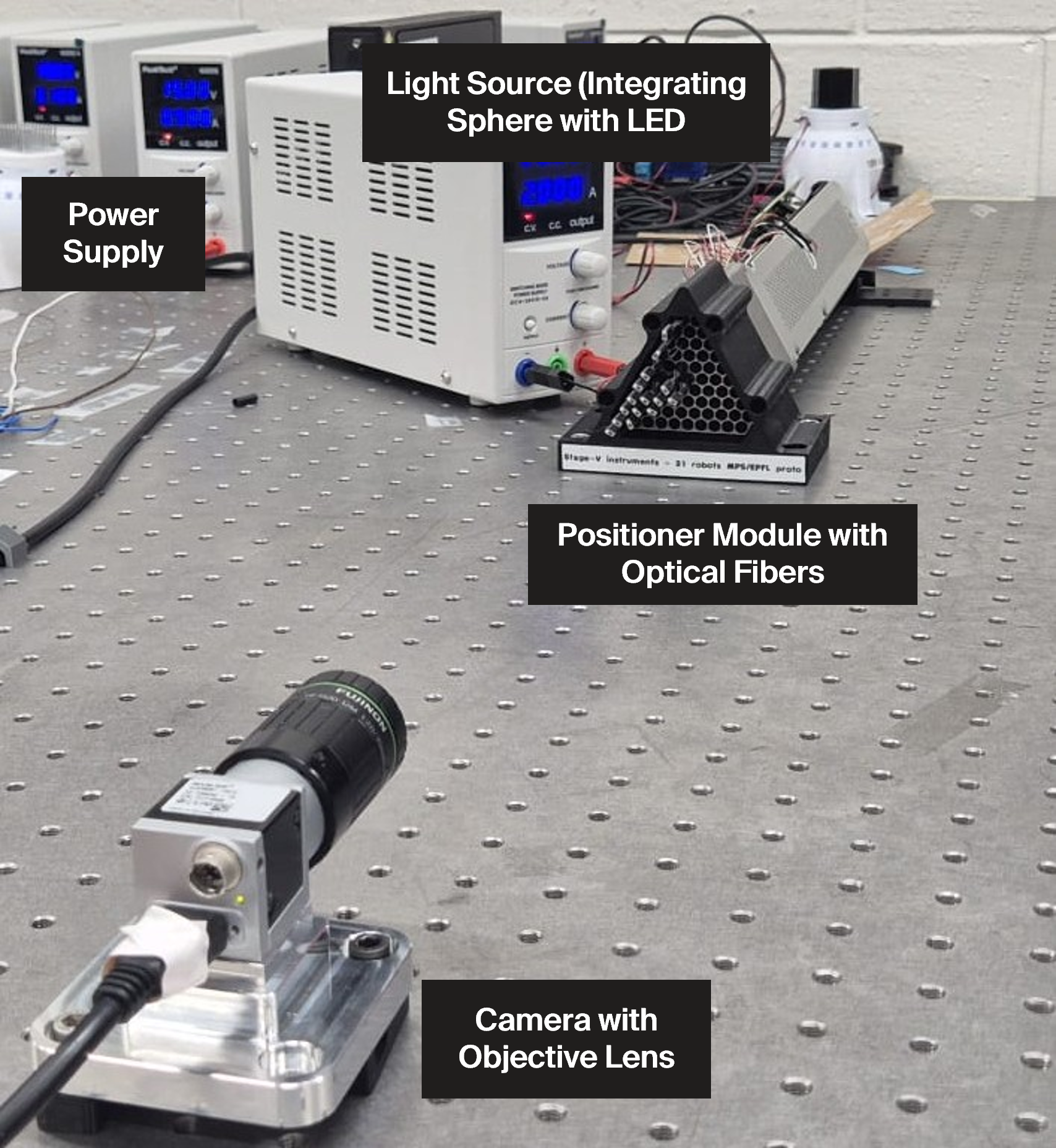}
    \vspace{.1cm}
    \caption{XY Positioning Performance Test Bench}
    \label{fig:setup}
\end{figure}

\begin{figure}[H]
  \makebox[\textwidth][c]{%
    \begin{minipage}{0.7\textwidth}
      \centering
      \captionsetup[subfigure]{justification=centering}
      \begin{subfigure}[b]{0.45\textwidth}
        \centering
        \includegraphics[width=\linewidth]{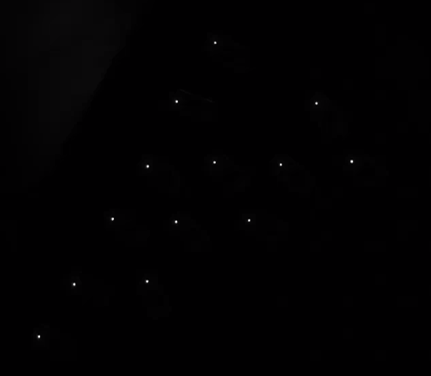}
        \caption{}
        \label{fig:multi_blob}
      \end{subfigure}
      \hspace{.3cm}
      \begin{subfigure}[b]{0.42\textwidth}
        \centering
        \includegraphics[width=\linewidth]{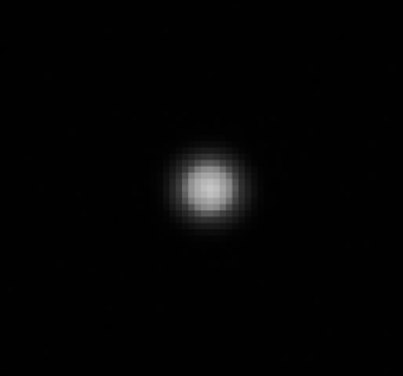}
        \caption{}
        \label{fig:blob_zoomedin}
      \end{subfigure}
    \end{minipage}
  }
  \vspace{.1cm}
  \caption{Illuminated spots captured by the camera; (a) image captured by the camera showing an array of backlit fibers; (b) Zoomed-in version of one of these spots}
  \label{fig:multi-spot}
\end{figure}

 The data presented in the figures below represent the mean root-mean-square (RMS) values obtained from five repetitions of the test program. The XY coordinates from the positioners in the module are determined by simultaneously acquiring multiple illuminated spots (Figure \ref{fig:multi-spot}) using a camera with an exposure time of 50 µs. The centroids of these spots are computed via two-dimensional Gaussian fitting. Figure \ref{fig:setup} shows the test-bench setup used.

As the orientation of the hard stops of the positioners in this prototype is mostly identical, it was possible to improve the data acquisition and command-handling software to send the same routine to all positioners simultaneously. Their movement was then detected concurrently, enabling parallelization of the test.

\subsection{ANGULAR TILT TEST}
\begin{figure}[H]
    \centering
    \includegraphics[width=0.7\linewidth]{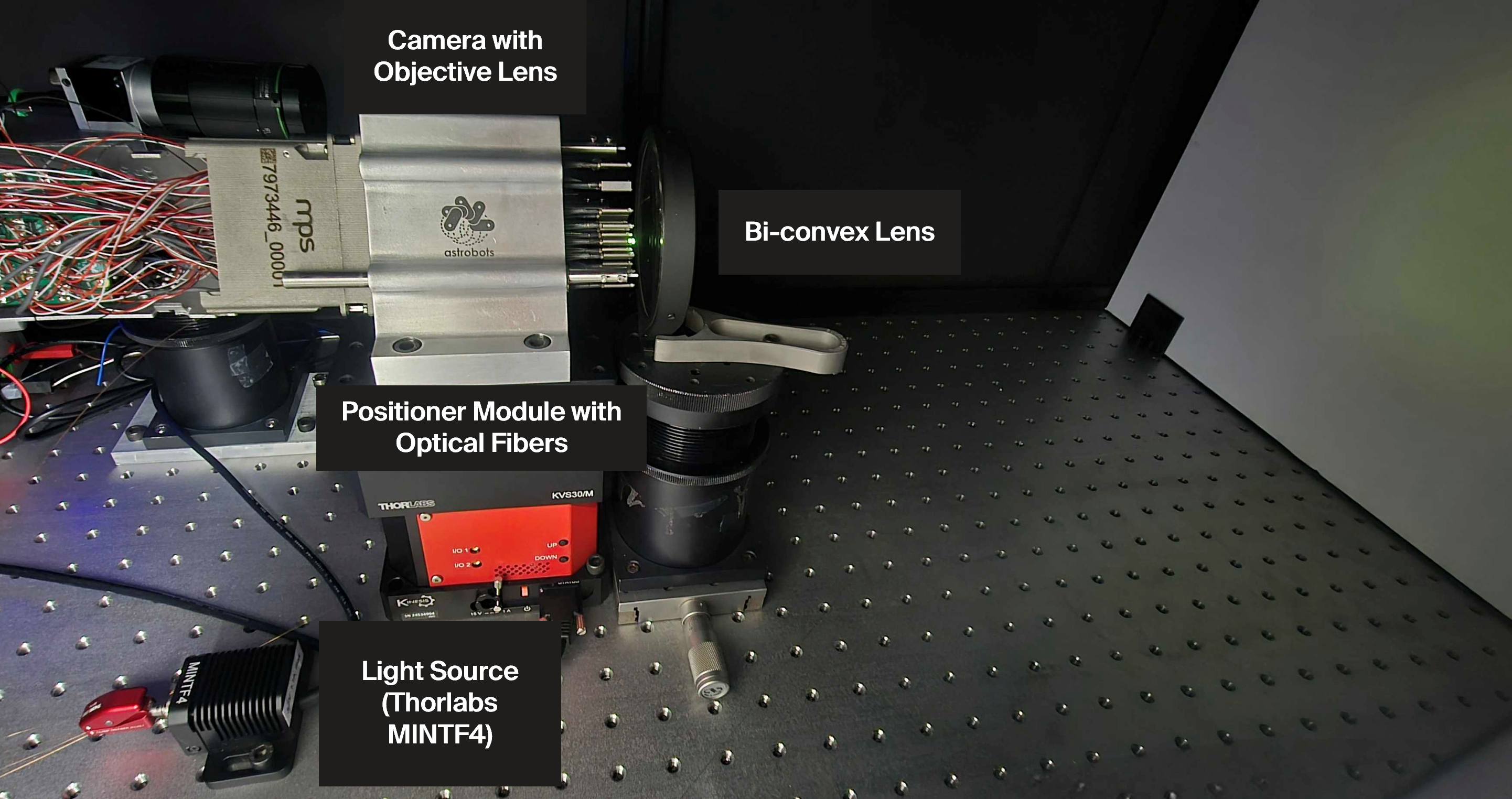}
    \vspace{.1cm}
    \caption{Angular Tilt Test Bench}
    \label{fig:setup2}
\end{figure}

Tilt tests are crucial to ensure that the positioners point to the desired orientation with as little deviation as possible. A tilt of maximum 0.4° is desired for each positioner. The individual tilt of the rotational axis of each robotic arm can be measured using the tilt test bench shown on figure \ref{fig:setup2}. By rotating each arm individually and looking at the displacement of the blob on a white screen placed at the focal length of the lens, we can disentangle xy displacement from tilt displacement \cite{10.1117/1.JATIS.6.1.018001}. Thus, we can deduce the tilt angle between the module axis and the alpha arm axis, between the beta arm axis and the alpha arm axis, and between the fiber axis and the beta arm axis. Details of the tilt test setup can be found in \citenum{Galal2025}.

\begin{figure}[H]
    \centering
    \includegraphics[width=0.7\linewidth]{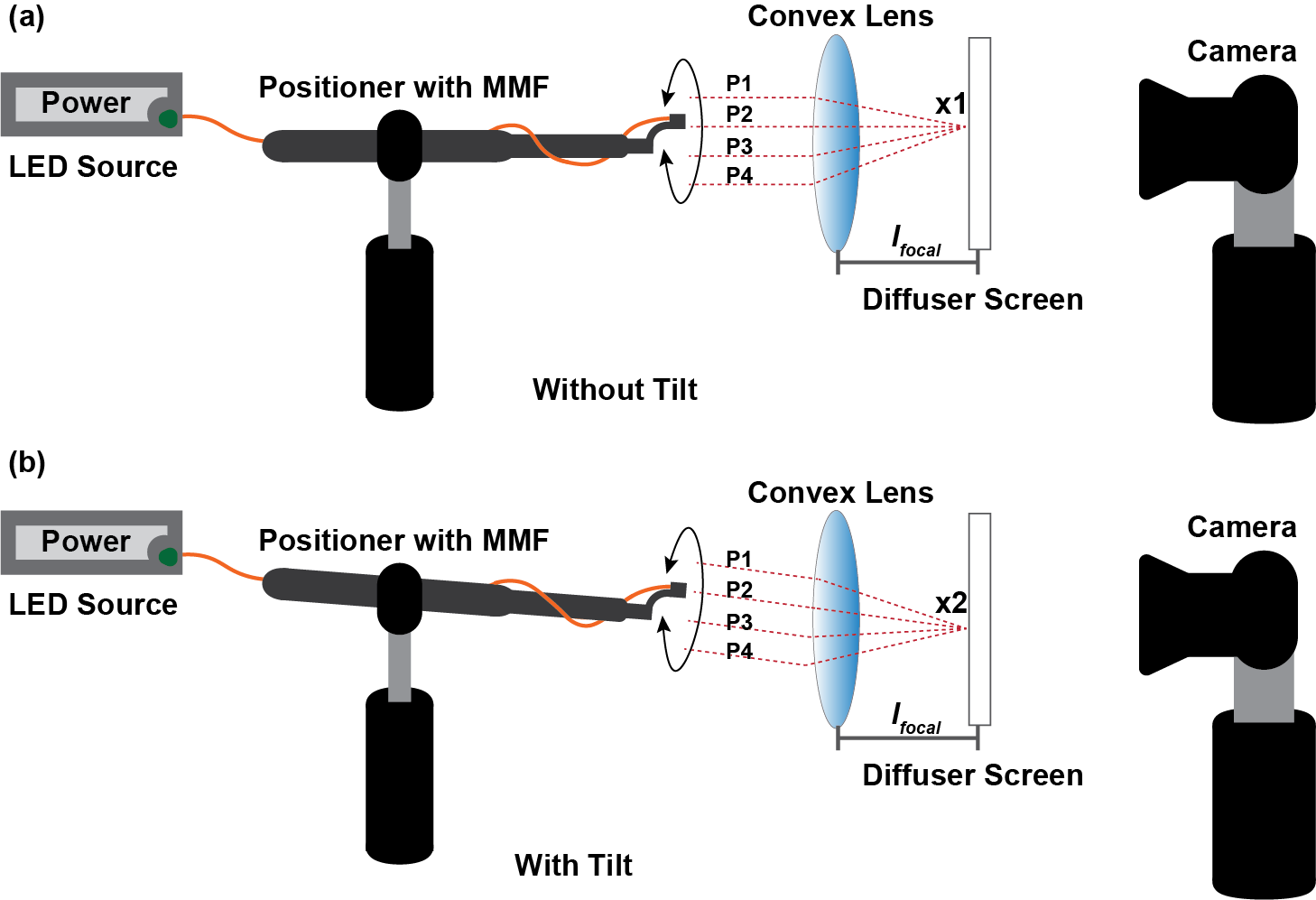}
    \vspace{.1cm}
    \caption{Principle of the Tilt Test \cite{Galal2025}}
    \label{fig:tilt_test_concept}
\end{figure}

\begin{figure}[H]
    \centering
    \includegraphics[width=0.5\linewidth]{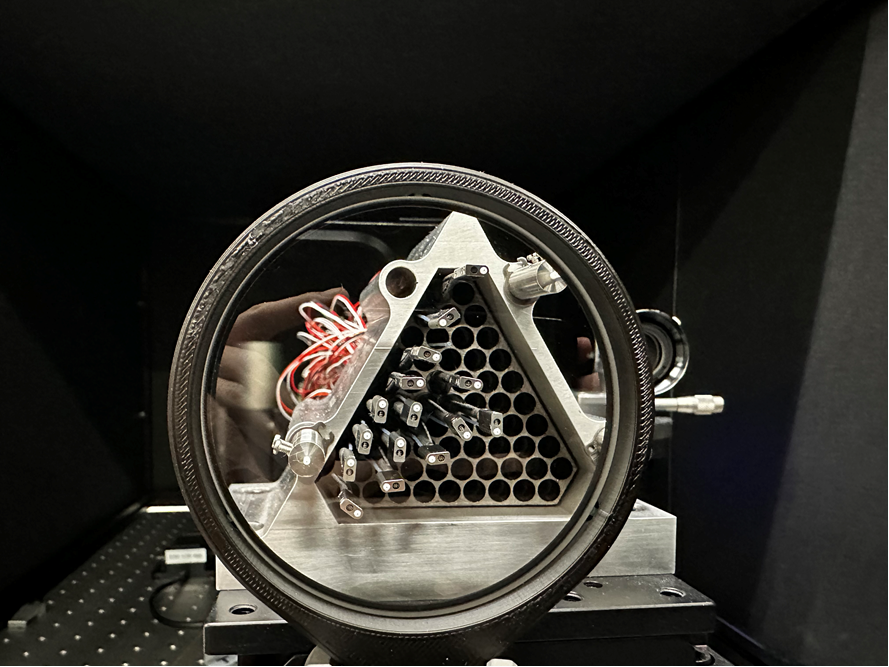}
    \vspace{.1cm}
    \caption{Calibration cylinder inside the module chassis support structure}
    \label{fig:calib_cylinder}
\end{figure}

On this prototype, we further refined the tilt test for the measurement of the tilt angle between the module axis and the alpha arm axis. As we cannot just remove a positioner from the chassis and put a calibration cylinder at the same location, an alternative method has been used. 
As can be seen on image \ref{fig:calib_cylinder}, the calibration cylinders are now placed on the side of the module. Assuming that the cylinders are aligned with the module axis, we can get the module axis position by rotating a chosen cylinder, and measuring the center of the drawn circle from the centroids of the projected light blobs on the screen. This center gives us the projected position of the module axis as seen on the white screen.

However, because there is a displacement from the cylinder position to the positioner position, and the disentanglement is not perfect, we also have to compensate for such displacement. This displacement is measured by moving the module with an illuminated fiber in the plane with an automated stage, and measuring the position of the centroid of the blob on the white screen. We obtained around 0.33 pixel of blob displacement on the screen for 1 mm of displacement of the illuminated fiber. The relative position between the cylinder and the chosen positioner is then used to compensate for this displacement in the post-processing.

Another observation on this method is that we would directly be measuring the tilt between the alpha axis and the module axis, assuming tight tolerances on the manufacturing of the chassis holder. Indeed, as the chassis is fixed to the holder through the 3 support faces, if the cylinders are straight with respect to the fixation faces of the chassis holder, then the module axis can be accurately determined.

\section{RESULTS}

The preliminary results presented in this Section give us an overview of the performance of the prototype. The results correspond to the mean values obtained from 5 repetitions of the test program.

\subsection{REPEATABILITY}

\begin{figure}[H]
    \centering
    \includegraphics[width=0.85\linewidth]{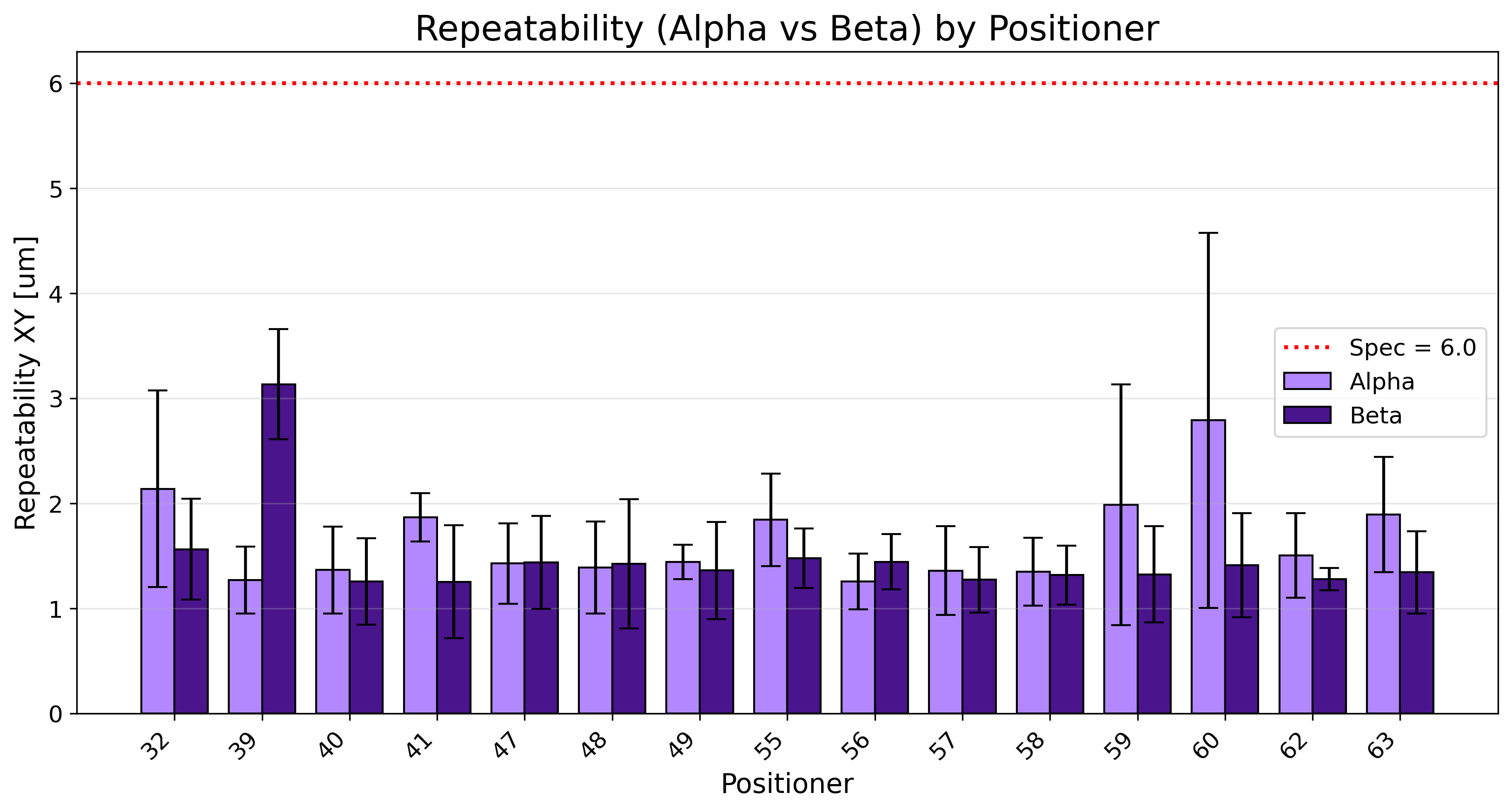}
    \vspace{.1cm}
    \caption{Repeatability performance per fiber-positioner. The bar plots show the root-mean-square results for the alpha and beta arm. The whiskers represent 1 standard deviation. Note: The red dotted line depicts the desired performance of \(\le \) 6 µm RMS for each arm.}
    \label{fig:rep_alpha_beta}
\end{figure}

The data used for calculating the repeatability of each positioner is acquired by commanding the alpha and beta arms repeatedly to a specific position. The data in Figure \ref{fig:rep_alpha_beta} shows the mean of the RMS values of the routine repeated 20 times for both alpha and beta arms. 

\subsection{DATUM REPEATABILITY}

\begin{figure}[H]
    \centering
    \includegraphics[width=0.85\linewidth]{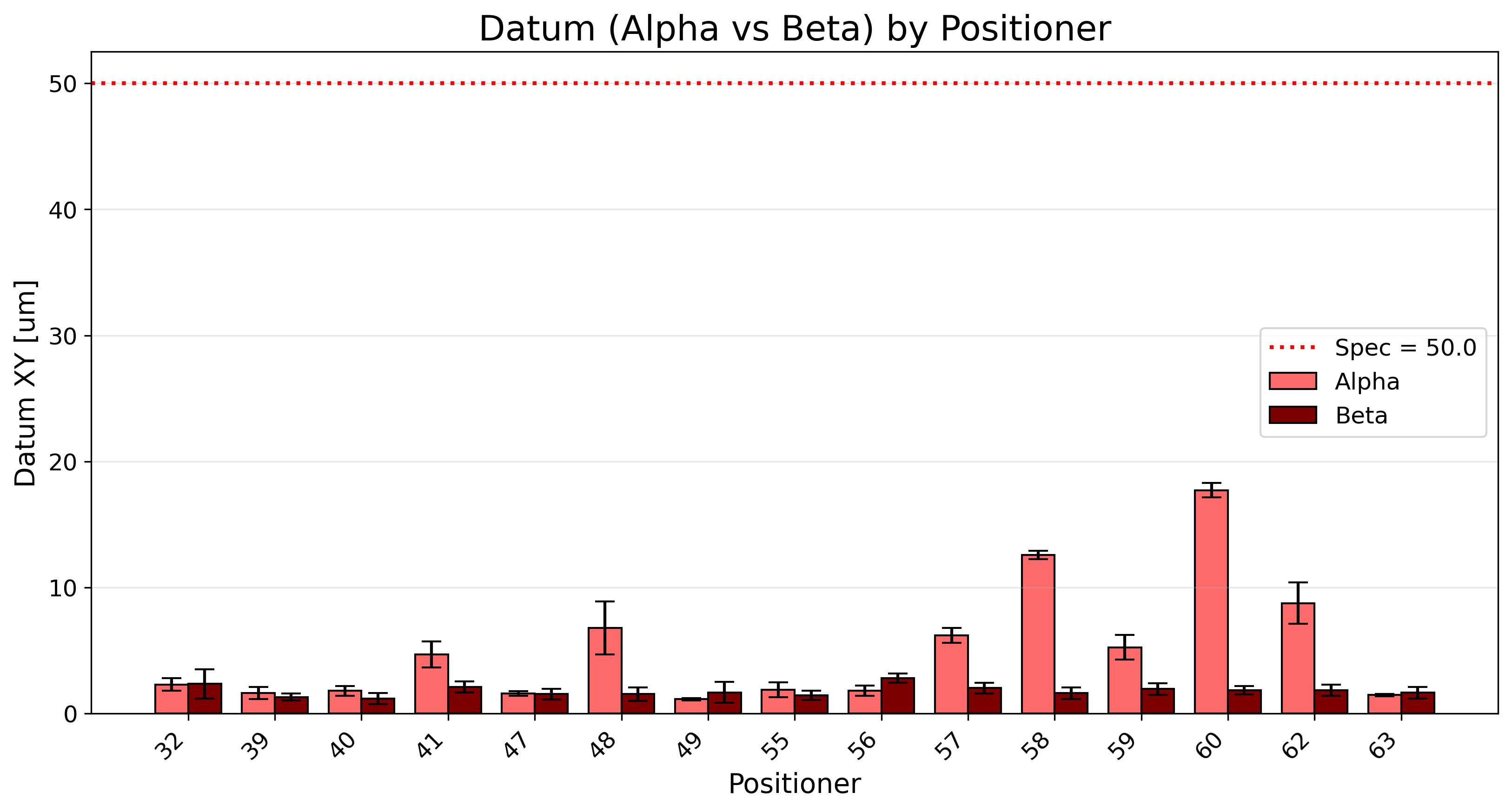}
    \vspace{.1cm}
    \caption{Datum repeatability performance per fiber-positioner. The bar plots show the root-mean-square results for the alpha and beta arm. The whiskers represent 1 standard deviation. Note: The red dotted line depicts the desired performance of \(\le \) 50 µm RMS for each arm.}
    \label{fig:datum_rep_alpha_beta}
\end{figure}

The data used for calculating the datum repeatability of each positioner is acquired by commanding the alpha and beta arms repeatedly to their hard stops. The data in Figure \ref{fig:datum_rep_alpha_beta} shows the mean of the values of the routine repeated 20 times for both alpha and beta arms. 

\subsection{BACKLASH}

In order to measure the backlash, the positioners are commanded to repeatedly move between two positions separated by a reference angular distance for 20 iterations and calculating the centroids of the spots for each position. The backlash is defined as the difference between the measured angular distance of the two centroids and the reference angular distance.

\begin{figure}[H]
    \centering
    \includegraphics[width=0.85\linewidth]{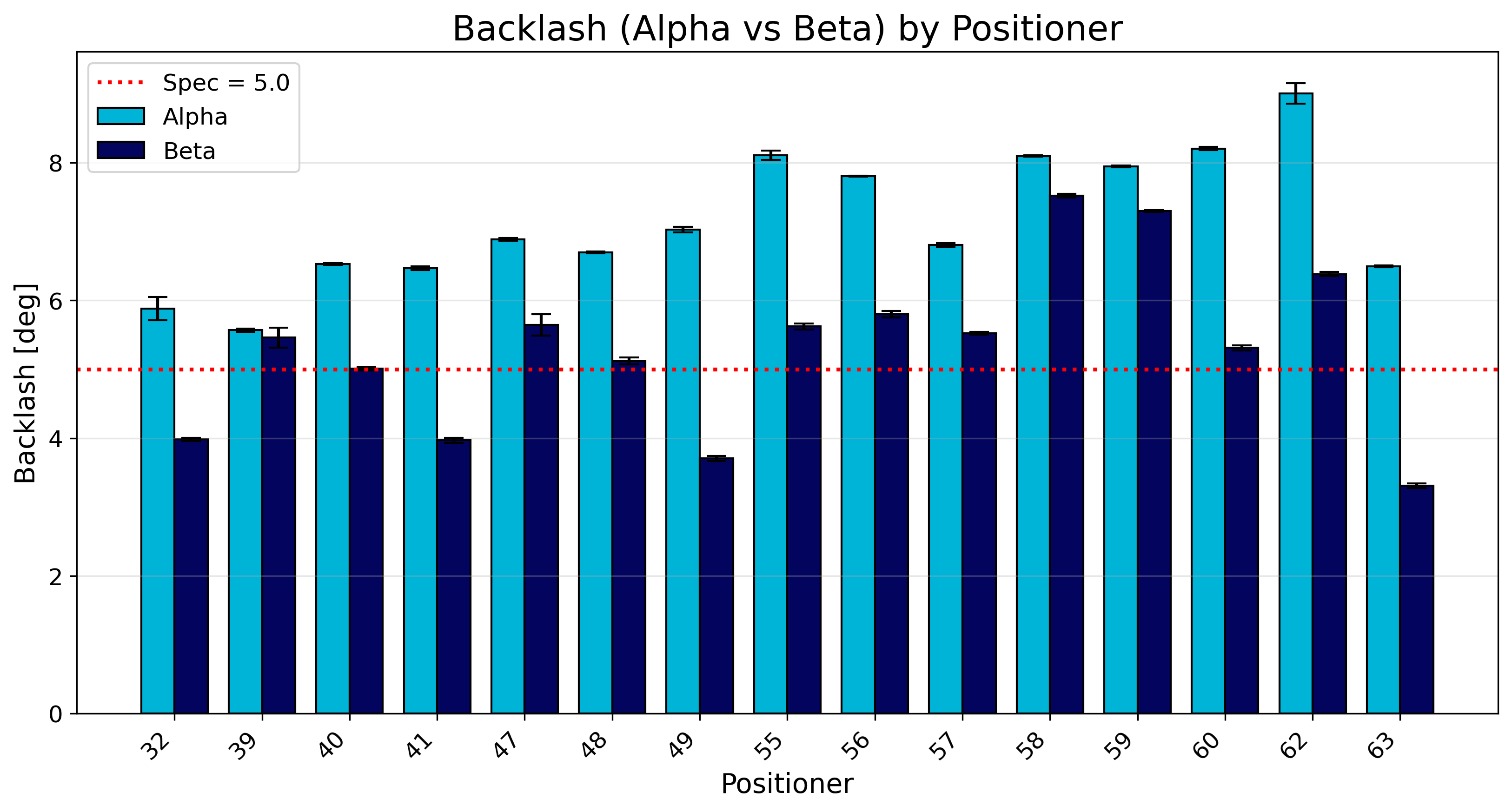}
    \vspace{.1cm}
    \caption{Calculated backlash per fiber-positioner. The bar plots show the obtained results for the alpha and beta arm. The whiskers represent 1 standard deviation. Note: The red dotted line depicts the desired performance of \(\leq 5^\circ\) for each arm.}
    \label{fig:backlash_alpha_beta}
\end{figure}

\subsection{MOTION RANGE}

The motion range of each positioner was tested by recording the movement from each one of its arms over the full angular range of both alpha and beta arms while performing steps of 1 degree. The motion range is measured using the same test as non-linearity. A summary of the results obtained is shown in Table \ref{tab:measmotionrange}.

\begin{table}[H]
    \centering
    \begin{tabularx}{0.55\textwidth}{p{0.7cm} Y Y Y Y}
    \toprule
        \textbf{Pos} & \textbf{Measured alpha motion range [°]} & \textbf{Measured beta motion range [°]} \\ \midrule
        32 & 369.85 & 190.15 \\ 
        39 & 370.56 & 192.70 \\ 
        40 & 370.61 & 190.61 \\ 
        41 & 371.56 & 189.49 \\ 
        47 & 370.18 & 189.52 \\ 
        48 & 370.29 & 191.92 \\
        49 & 369.41 & 190.62 \\ 
        55 & 369.11 & 190.13 \\ 
        56 & 368.80 & 190.44 \\ 
        57 & 370.82 & 191.66 \\ 
        58 & 369.16 & 189.48 \\ 
        59 & 369.78 & 189.93 \\ 
        60 & 370.68 & 191.69 \\ 
        62 & 367.81 & 190.80 \\ 
        63 & 371.31 & 190.82 \\ \bottomrule
    \end{tabularx}
    \vspace{1em}
    \caption{Motion range summary}
    \label{tab:measmotionrange}
\end{table}

\begin{figure}[H]
  \makebox[\textwidth][c]{%
    \begin{minipage}{1\textwidth}
      \centering
      \captionsetup[subfigure]{justification=centering}
      \begin{subfigure}[b]{0.495\textwidth}
        \centering
        \includegraphics[width=\linewidth]{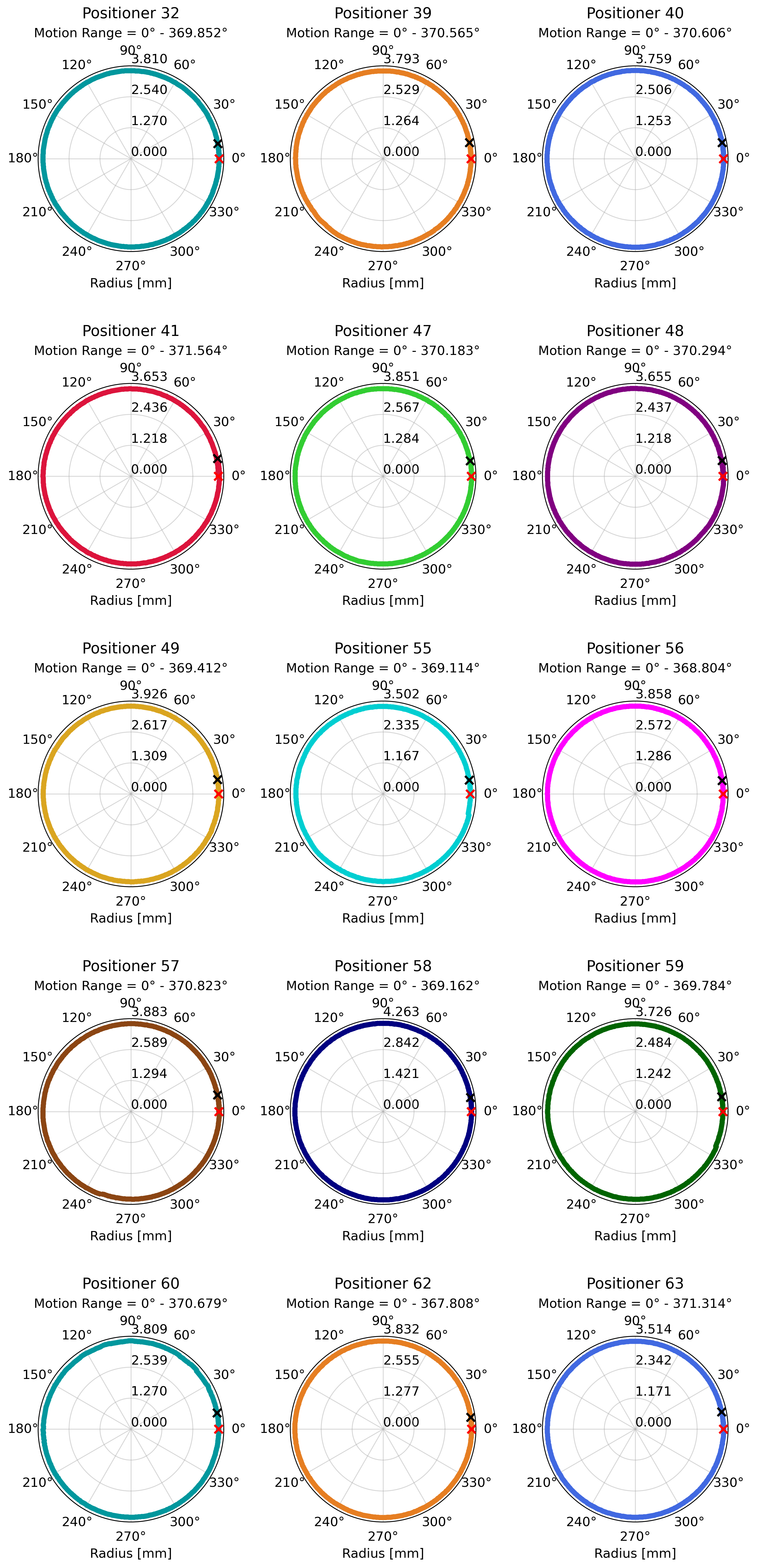}
        \caption{}
        \label{fig:motionrangealpha}
      \end{subfigure}
      \begin{subfigure}[b]{0.495\textwidth}
        \centering
        \includegraphics[width=\linewidth]{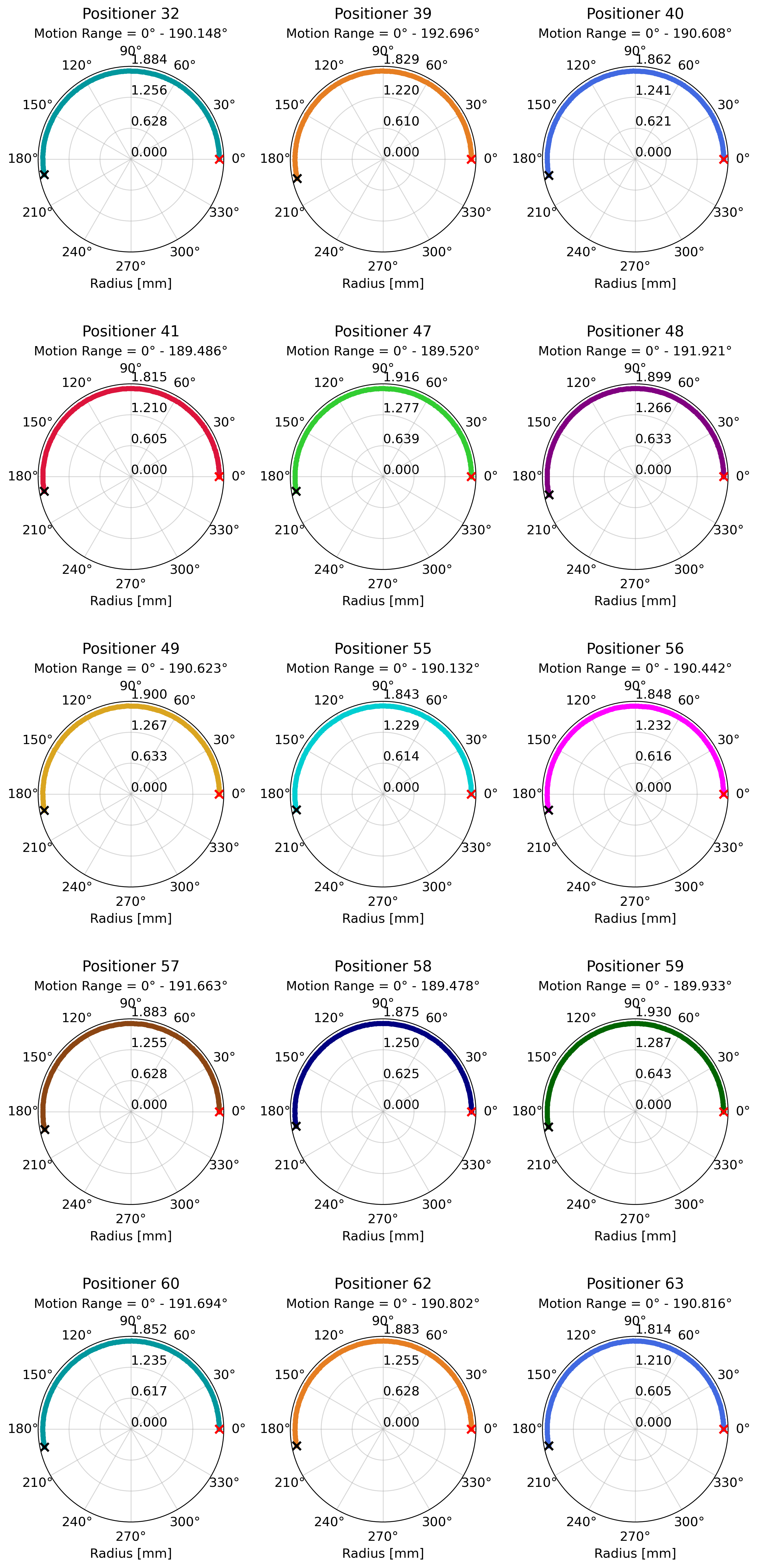}
        \caption{}
        \label{fig:motionrangebeta}
      \end{subfigure}
    \end{minipage}
  }
  \vspace{.1cm}
  \caption{Motion range per fiber-positioner for alpha (a) and beta arms (b). Note: Red 'x' is the start, and black 'x' is the end. Alpha arm desired performance = 370°; Beta desired performance = 190°.}
  \label{fig:motionrange}
\end{figure}

Figure \ref{fig:motionrange} shows the radial plot of the movement throughout the motion range of both alpha and beta arms. As observed the reconstructed data depicts a circle for the alpha arm and an arc for the beta arm. The radial axis of the alpha arm plots from Subfigure \ref{fig:motionrangealpha} show the measured distance from the center of each positioner's workspace to the ferrule's center when both arms are fully extended. Thus being equivalent to the sum of both arms. The radial axis of the beta arm is equivalent to the beta arm length. It is important to note that the measured arm length here are influenced by the tilt of the beta arm and the tilt of the ferrule, they are thus not to be taken as accurate measurements of the physical arm lengths.

\subsection{NON LINEARITY}

\begin{figure}[H]
  \makebox[\textwidth][c]{%
    \begin{minipage}{1\textwidth}
      \centering
      \captionsetup[subfigure]{justification=centering}
      \begin{subfigure}[b]{0.495\textwidth}
        \centering
        \includegraphics[width=\linewidth]{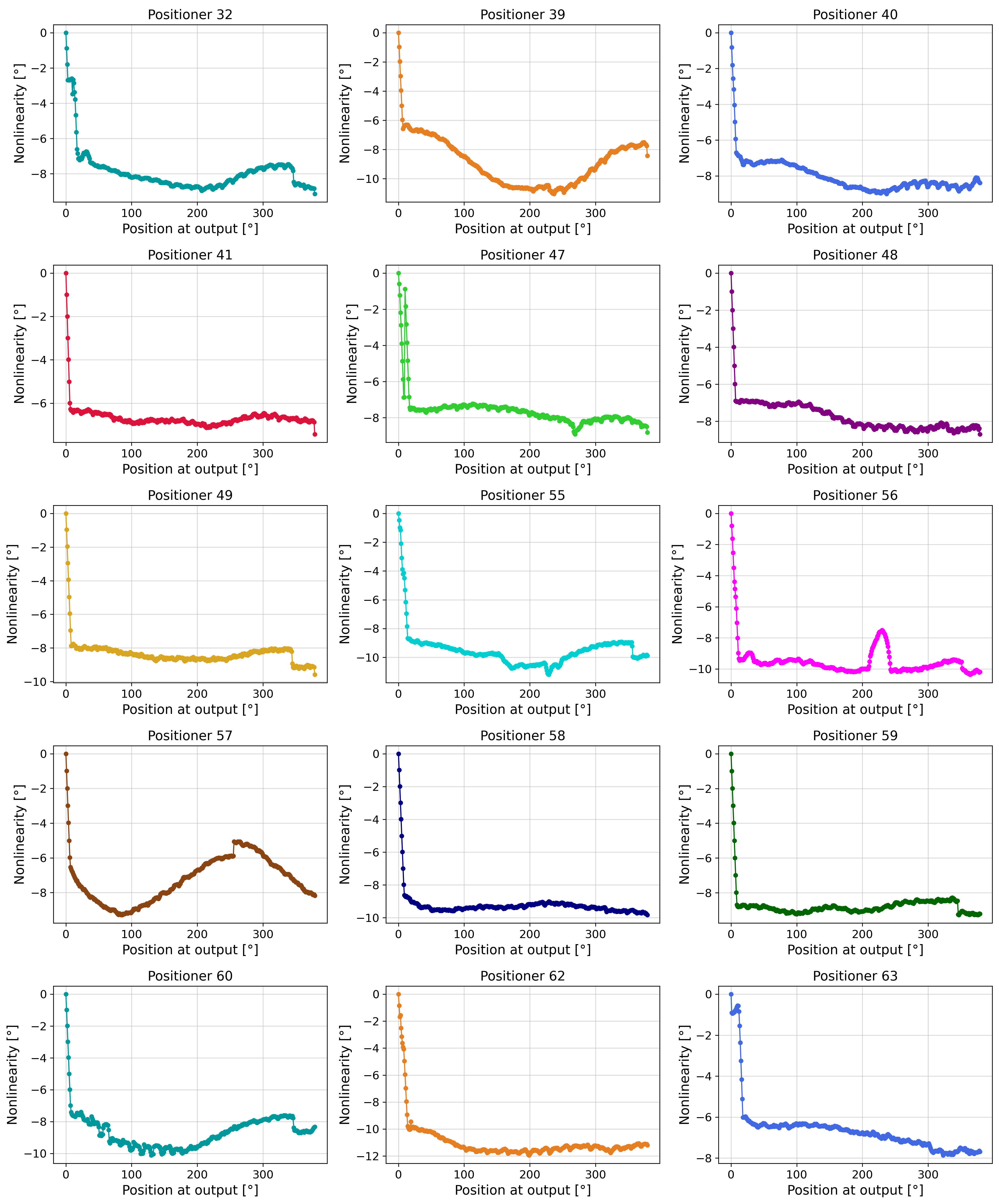}
        \caption{}
        \label{fig:nlalpha}
      \end{subfigure}
      \begin{subfigure}[b]{0.495\textwidth}
        \centering
        \includegraphics[width=\linewidth]{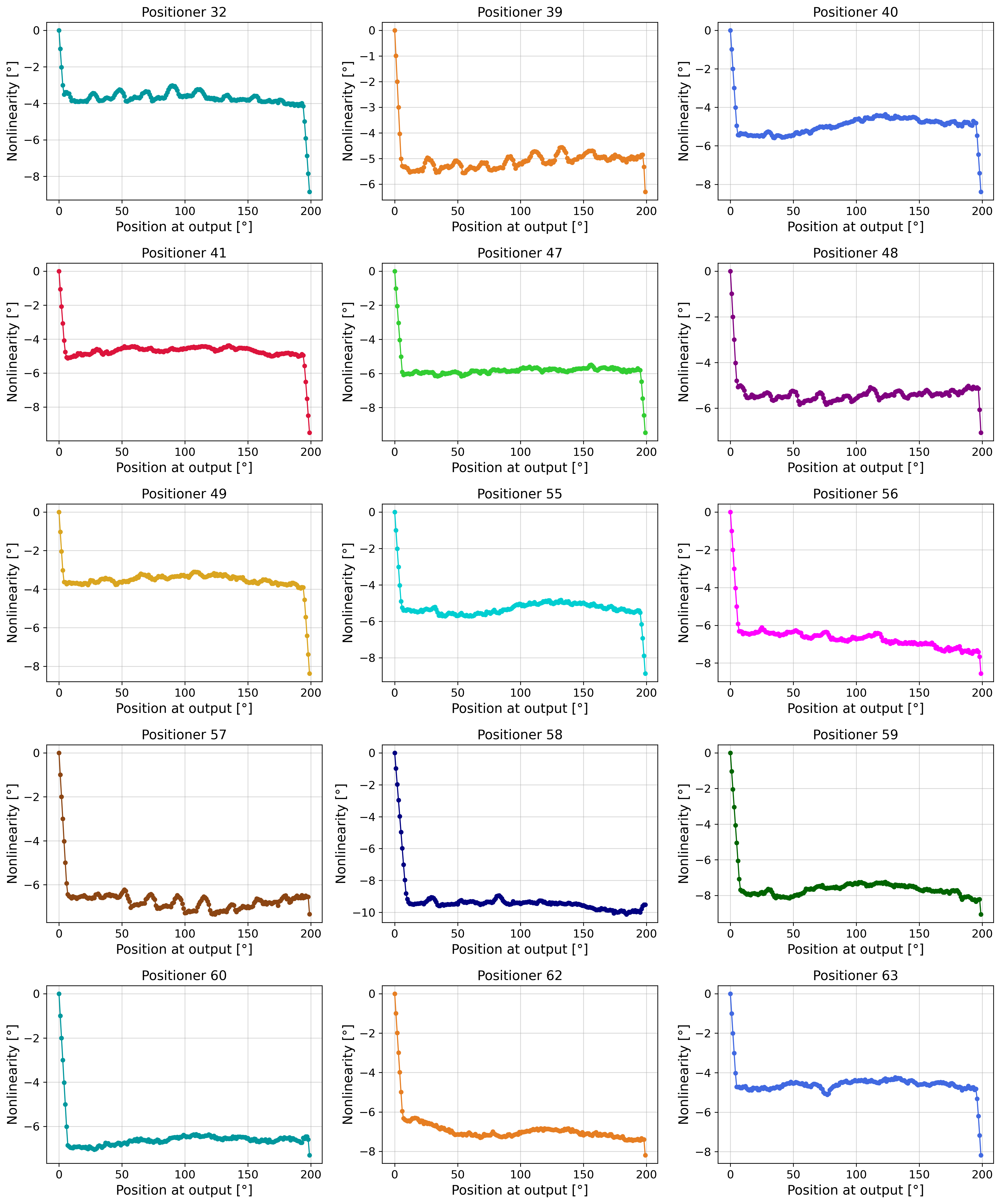}
        \caption{}
        \label{fig:nlbeta}
      \end{subfigure}
    \end{minipage}
  }
  \vspace{.1cm}
  \caption{Non-linearity graphs for (a) alpha and (b) beta arms of each fiber-positioner.}
  \label{fig:nl}
\end{figure}

Using the data acquired to reconstruct the motion range of each positioner we can also calculate their non-linearity. As long as the non-linearity behavior is repeatable, the effect can normally be compensated for. Figure \ref{fig:nl} shows examples of the non-linear behavior obtained in an iteration for the alpha and beta arms of each positioner.

As stated in Ref. \citenum{kronig_precision_2020} and observed in Ref. \citenum{Galal2025} the high-frequency variations in transmission error are caused by deviations from a perfect involute tooth profile. These imperfections make the line of action slightly curved instead of straight, producing rapid, repeating fluctuations that are likely tied to the number of teeth in the gear. In contrast, the low-frequency variations arise from larger-scale geometric or assembly issues, such as an imperfect base circle. These can be due to run-out, eccentricity, or misalignment during mounting, leading to slower, more gradual changes in the transmission error.

\subsection{ARM LENGTH}\label{subsec:armlength}

\begin{figure}[H]
  \makebox[\textwidth][c]{%
    \begin{minipage}{1\textwidth}
      \centering
      \captionsetup[subfigure]{justification=centering}
      \begin{subfigure}[b]{0.495\textwidth}
        \centering
        \includegraphics[width=\linewidth]{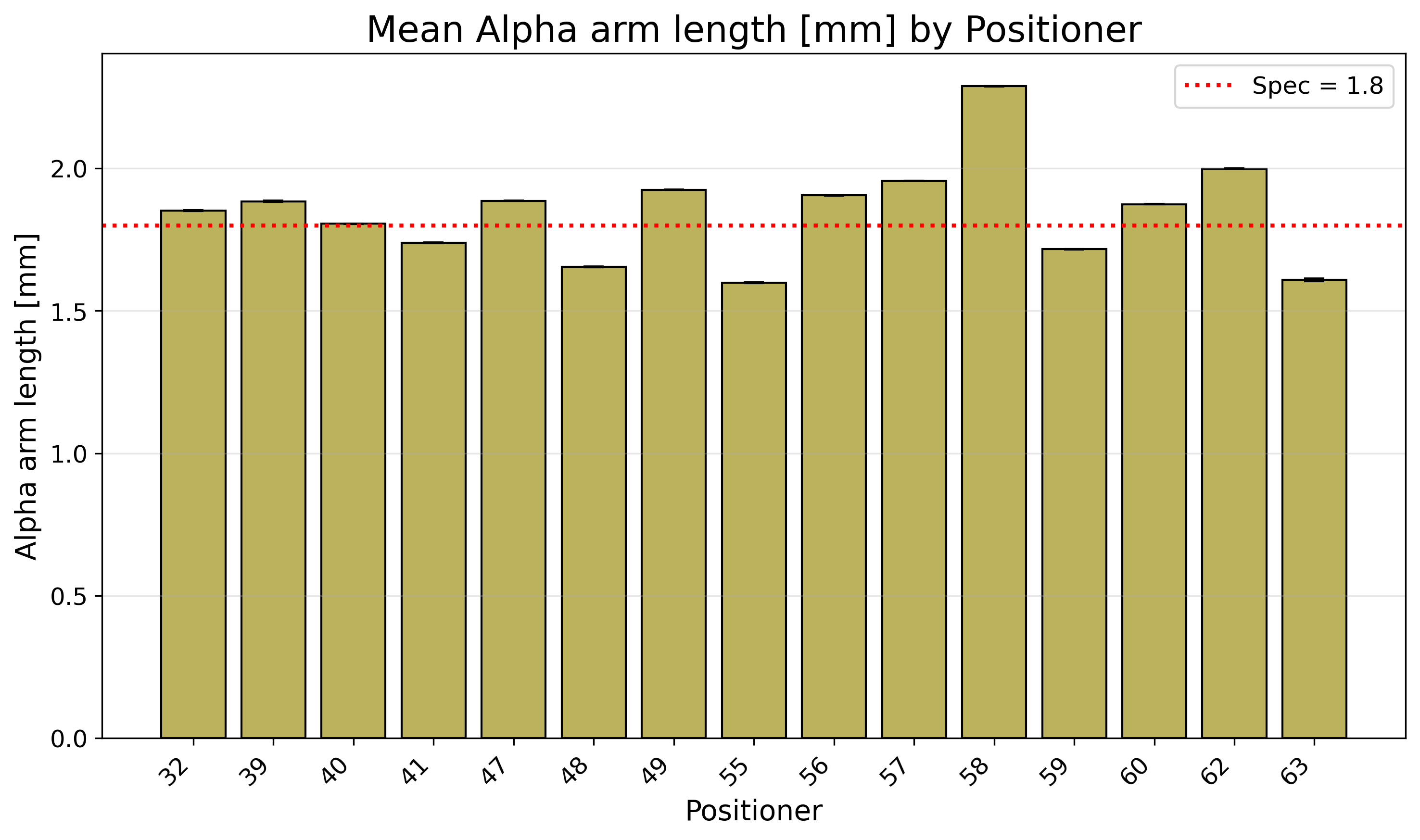}
        \caption{}
        \label{fig:armlengthalpha}
      \end{subfigure}
      \begin{subfigure}[b]{0.495\textwidth}
        \centering
        \includegraphics[width=\linewidth]{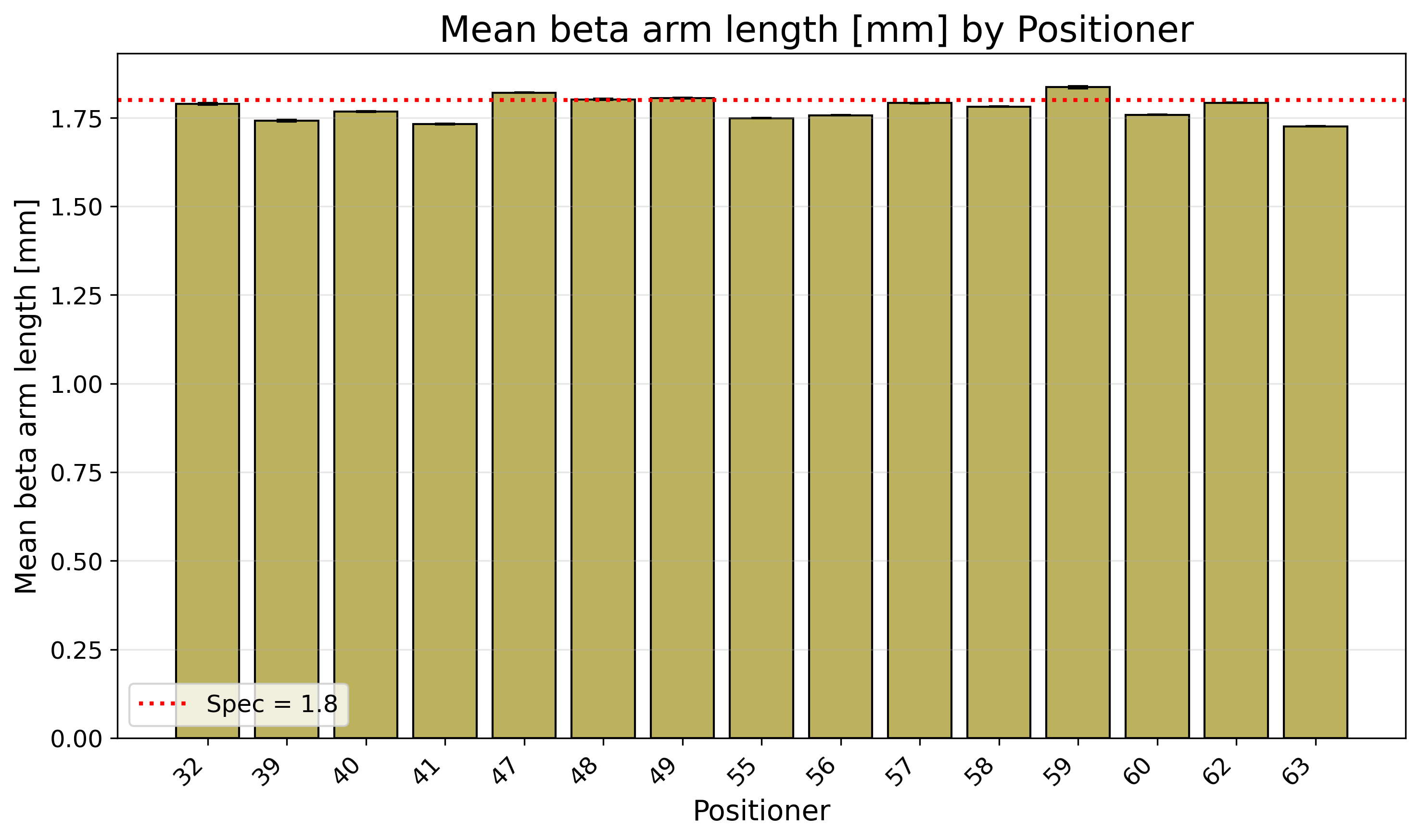}
        \caption{}
        \label{fig:armlengthbeta}
      \end{subfigure}
    \end{minipage}
  }
  \vspace{.1cm}
  \caption{Calculated arm length per fiber-positioner. The bar plots show the root-mean-square values versus the positioner number; (a) results for the alpha arm, and (b) results for the beta arm. Note: The red dotted line depicts the desired length of 1.8 mm for each arm.}
  \label{fig:armlength}
\end{figure}

The arm lengths are determined by analyzing the centroid from the trajectories commanded as each positioner arm is moved independently through its full motion range. The beta arm length is obtained directly from the radius of a circle fitted to the measured beta arc points. The alpha arm length is estimated by first fitting beta arcs at several alpha positions, then extracting the centers of those fitted circles and performing a second circle fit to those centroids. It is important to note that the measured arm lengths are not the physical arm lengths of the positioners, but the "optical" arm length as seen by the camera. This optical arm length is influenced not only by the mechanical length of the arms, but also by the tilt of the different arms, the tilt of the ferrule within the beta arm and the tilt of the fiber within its ferrule. The resulting arm length are presented in Figure \ref{fig:armlength}.

\subsection{TILT TEST}

During the tilt test, each arm is moved individually over their whole motion range, and that at different positions of the other arm. At each step, the centroid of the light spot on the screen is captured. The center of the fitted circle made of those centroids is then taken for each full motion range movement for each arm. The tilt between the alpha arm and the reference ($\theta$) is taken as the distance from the center of the alpha circle and the circle drawn from the cylinder. The tilt between the beta arm and the alpha arm ($\phi$) is taken as the converted distance from the center of the beta circle and the alpha circle. The RMS between measurements are then taken. Details about the tilt can be found in \citenum{10.1117/1.JATIS.6.1.018001} and \citenum{Galal2025}.

\begin{table}[H]
    \setlength{\tabcolsep}{0.2pt}
    \centering
    
    \begin{tabularx}{\textwidth}{@{}p{0.7cm} *{7}{Y}@{}}
    \toprule
        \textbf{Pos} & \textbf{Tilt $\theta$ [°]} & \textbf{$\theta$ std} & \textbf{Tilt $\phi$ [°]} & \textbf{$\phi$ std} & \textbf{Direct Sum [°]} & \textbf{RSS [°]} & \textbf{P/NP}\\ \midrule
        32 & 0.29 & 0.049 & 0.13 & 0.034 & 0.42 & 0.32 & P\\ 
        39 & 0.25 & 0.009 & 0.20 & 0.013 & 0.45 & 0.32 & P\\ 
        40 & 0.29 & 0.019 & 0.51 & 0.026 & 0.80 & 0.59 & NP\\ 
        41 & 0.18 & 0.021 & 0.14 & 0.025 & 0.32 & 0.23 & P\\ 
        47 & 0.51 & 0.018 & 0.24 & 0.75 & 0.022 & 0.56 & NP\\ 
        48 & 0.32 & 0.022 & 0.52 & 0.015 & 0.84 & 0.61 & NP\\  
        49 & 0.30 & 0.008 & 0.19 & 0.018 & 0.49 & 0.36 & P\\ 
        55 & 0.68 & 0.007 & 0.35 & 0.018 & 1.03 & 0.76 & NP\\ 
        56 & 0.29 & 0.032 & 0.26 & 0.046 & 0.55 & 0.39 & P\\ 
        57 & 0.36 & 0.006 & 0.22 & 0.010 & 0.58 & 0.42 & NP\\ 
        58 & 0.28 & 0.028 & 0.58 & 0.018 & 0.86 & 0.65 & NP\\ 
        59 & 0.21 & 0.008 & 0.17 & 0.02 & 0.38 & 0.26 & P\\ 
        60 & 0.31 & 0.007 & 0.1 & 0.012 & 0.41 & 0.33 & P\\
        62 & 0.18 & 0.018 & 0.61 & 0.032 & 0.79 & 0.63 & NP\\ 
        63 & 0.29 & 0.007 & 0.47 & 0.016 & 0.76 & 0.55 & NP \\\bottomrule
    \end{tabularx}
    \vspace{1em}
    \caption{Tilt testing results per fiber positioner. The desired performance is \(\le \) 0.4°. P/NP (pass/not passed) indicates whether the RSS is over or below 0.4°. The reported tilts are rms values. $\theta$ is the tilt of the alpha arm and $\phi$ is the tilt of the beta arm.}
\end{table}

\subsection{SUMMARY}

The performance results from this prototype when compared to desired performances for Stage-5 telescope projects is shown in Table \ref{table: comparison}.

\begin{table}[H]
    \centering

    \begin{tabularx}{0.8\textwidth}{p{5.5cm} Y Y Y Y}
        \toprule
        \textbf{Parameter} & 
        \textbf{Desired Perf.} & 
        \textbf{Best Unit} & 
        \textbf{Worst Unit} & 
        \textbf{Average}\\
        \midrule\midrule
        
        Repeatability $\alpha$ ($\mu m$ RMS) & $\leq$6 & 1.26 & 2.79 & 1.66\\
        Repeatability $\beta$ ($\mu m$ RMS) & $\leq$6 & 1.25 & 3.14 & 1.49  \\
        Repeatability (Comb.) ($\mu m$ RMS) & -- & 1.78  & 4.20 & 2.23 \\
        
        \midrule
        
        Datum Rep. $\alpha$ ($\mu m$ RMS)  & $\leq$50 & 1.14 & 17.72 & 5.04\\
        Datum Rep. $\beta$ ($\mu m$ RMS)   & $\leq$50 & 1.18 & 2.82 & 1.80\\
        Datum Rep. (Comb.) ($\mu m$ RMS)   & -- & 1.64 & 17.95 & 5.35\\
        
        \midrule
        
        Backlash $\alpha$ (deg) & $\leq$5 & 5.57 & 9.01 & 7.17\\
        Backlash $\beta$ (deg) & $\leq$5 & 3.31 & 7.52 & 5.31\\
        
        \midrule
        
        Angular Tilt (deg) & $\leq$0.4 & 0.23 & 0.76 & 0.47\\
        
        \bottomrule
    \end{tabularx}
    \vspace{1em}
    \caption{Comparison matrix showing the results of the MPS prototype and the desired performances for Stage-5 instruments. Results are shown for the best and worst performing fiber-positioner units as well as the average from the module. Note: Combined values for Repeatability and Small Move Error are calculated as the quadrature sum of the $\alpha$ and $\beta$ components.}
    \label{table: comparison}
\end{table}

\section{DISCUSSION}
The test results are overall great for this 2nd iteration of the MPS prototype. Repeatability is below the desired 6 [um] for all tested positioners, with most positioners being even below 2 [um]. The average datum repeatability is also significantly lower than the desired 50 [um] with most positioners being below 10 [um]. Repeatability measurement results are radial rms values.

The measured backlash of most positioners is however higher than the desired performance of 5°. This value has been fixed as to simplify the collision avoidance problem when moving the positioners. As we choose to approach all the targets from the same rotational direction for each arm in order to compensate the backlash, we need to keep the maximum backlash value as low as possible to reduce the range of that correction movement. Further investigation with the collision avoidance algorithm is necessary in order to understand better the impact of having higher values of backlash.

The measured motion ranges come very close or exceed the desired performances of 370° for alpha and 190° for beta. While only 8/15 alpha motion ranges and 11/15 beta motion ranges exceed the desired performance, the other motion ranges come very close to those values. However, it is important to note that those values have been defined while the desired backlash is at 5°. As a significant number of positioners exceed that value, the margin for the same direction approach is reduced and could lead to not being able to reach some edge positions within the patrol area of the positioner. Indeed, the measured motion range minus the nominal motion range (360° for alpha and 180° for beta) should at at least exceed the measured backlash for that same positioner: the positioner would need that extra range in order to perform a same side approach at the end of its range.

Plots in \ref{fig:nl} show the nonlinear behaviors of each positioner. The more drastic changes on the edges are due to backlash and when the positioner has reached its hardstop (no observed motion after commanded motion). Some positioners show low frequency variations, mostly due to mechanical imperfections. What is desired here is a smooth line with as little sudden jumps as possible. Overall, nonlinear behaviors look regular enough for most positioners within their motion range. In order to confirm whether those nonlinearities are acceptable for the desired accuracy they have to reach (5 µm), accuracy tests incorporating those values would have to be performed.

The obtained arm length values can differ significantly from the nominal arm lengths. Knowing the tight tolerances on the manufacturing and assembly of the positioners, the physical arm length cannot differ as much from the nominal value. As stated in \ref{subsec:armlength}, the optical arm length is influenced by the tilt of the different parts of the positioner. Still, those measurements are necessary in order to calibrate each positioner.

Finally, we observe that 7/15 positioners have a root sum square (RSS) tilt under the desired 0.4°. Direct sum is representative of a worst case scenario, where both alpha and beta tilts are aligned. Generally speaking, both tilts will not be perfectly aligned, so the RSS would be a more representative value of a typical tilt on any given observation. The standard deviations of the tilt of both axes are also reported, as they also give us information of whether the rotational axes themselves move in different positions. But we can observe that they are very low, suggesting that the axes do not move.

\section{CONCLUSION}
Current MPS prototype testing results show excellent repeatability and datum repeatability performances. However, average backlash values at 7.17° for alpha and 5.31° for beta are mostly higher than the desired 5°. The average tilt of 0.47° is also slightly higher than the desired 0.4°. Overall, performances are great and the MPS 15-positioner prototype makes a significant step towards a modular high density fiber positioner solution. Further tests such as accuracy, thermal \cite{rombach2026thermal} and lifetime test will be performed on this prototype. The accuracy test will be especially important in order to evaluate if the current prototype can reach the desired accuracy, and within how many moves for each positioner. As a next step, similar tests will also be performed on the next MPS prototype of 63 positioners to compare performances. Other tests such as focal ratio degradation, throughput tests and orientation tests \cite{wuthrich2026gravity} are also planned for current and future prototypes.

\acknowledgments 
The authors would like to acknowledge Innosuisse (Funding No.: 101.014 IP-ENG) for supporting
this work. Additionally, the authors would like to thank Dr. Luzius Kronig for the fruitful discussions and his tremendous help. Finally, the authors are very thankful to EPFL workshops for their
expertise in making precision parts for the test-benches.

\bibliography{report} 
\bibliographystyle{spiebib} 

\end{document}